\RequirePackage{fix-cm}
\documentclass[twocolumn,epjc3]{svjour3}
\smartqed  
\usepackage{amsmath}
\usepackage{amssymb}
\usepackage{amssymb,bbold}
\usepackage{kantlipsum,widetext}

\RequirePackage{graphicx}
\RequirePackage[colorlinks,citecolor=blue,urlcolor=blue,linkcolor=blue]{hyperref}
\RequirePackage[numbers,sort&compress]{natbib}

\journalname{Eur. Phys. J. C}
\begin{document}

\title{Comment on ``Dirac fermions in Som-Raychaudhuri space-time with scalar and vector potential and the energy momentum distributions [Eur. Phys. J. C (2019) 79: 541]''
}


\author{Francisco A. Cruz Neto\thanksref{addr1}\and Andr\'{e}s G. Jir\'{o}n Vicente\thanksref{addr1,addr3} \and Luis B. Castro\thanksref{e1,addr1,addr2} 
}

\thankstext{e1}{e-mail: luis.castro@ufma.br, luis.castro@pq.cnpq.br}


\institute{Departamento de F\'{\i}sica, Universidade Federal do Maranh\~{a}o (UFMA), Campus Universit\'{a}rio do Bacanga, 65080-805, S\~{a}o Lu\'{\i}s, MA, Brazil.
\label{addr1} \and
Universidad Tecnol\'{o}gica del Per\'{u} (UTP), Av. N. Ayllon Km 11.60, Lt 96, ATE, Lima, Per\'{u}.
\label{addr3} \and
Departamento de F\'{\i}sica e Qu\'{\i}mica, Universidade Estadual Paulista (UNESP), Campus de Guaratin\-gue\-t\'{a}, 12516-410, Guaratinguet\'{a}, SP, Brazil.\label{addr2}
}

\date{Received: date / Accepted: date}

\maketitle


In a recent paper in this Journal, P. Sedaghatnia et. al. \cite{EPJC79:541:2019} have studied the Dirac equation in presence of scalar and vector potentials in a class of flat G\"{o}del-type space-time called Som-Raychaudhuri space-time by using the methods quasi-exactly solvable (QES) differential equations and the Nikiforov Uvarov (NU) form. To achieve their goal, the authors have mapped the system into second-order differential equation (Schr\"{o}dinger-like problem). It is worthwhile to mention that expressions obtained [Eqs. (2.8)-(2.17)] in Ref. \cite{EPJC79:541:2019} are correct. On the other hand, the second order differential equation in Ref. \cite{EPJC79:541:2019} is wrong, probably due to erroneous calculations in the manipulation of the two coupled first order differential equations. This fact jeopardizes the results of \cite{EPJC79:541:2019}. The purpose of this comment is to calculate the correct differential equation and following the appropriate procedure obtain the solution for this problem.

The G\"{o}del-type solution with torsion and a topological defect can be written in cylindrical coordinates by the line element
\begin{equation}
ds^2=-(dt+\alpha \Omega r^2 d\varphi)^2+dr^2+\alpha^2r^2d\varphi^2+dz^2\,. \label{1}
\end{equation}
\noindent The Dirac equation for a free Fermi field $\Psi$ of mass $M$ in a Som-Raychaudhuri space-time with scalar and vector potentials is given by \cite{EPJC79:541:2019}
\begin{equation}
    \left[i\gamma^\mu\left(\nabla_\mu+ieA_\mu\right)-\left(M+S(r)\right)\right]\Psi(r,t)=0  \label{2}
\end{equation}
\noindent where $A_\mu=(V(r),0,0,0)$, $\nabla_\mu=(\partial_\mu+\Gamma_\mu)$ and $\Gamma_\mu$ is the affine connection. Now, we using the correct results of Ref. \cite{EPJC79:541:2019} and considering the solution in the form 
\begin{equation}
    \Psi(t,r,\varphi,z)=e^{-iEt}e^{i(m\varphi+kz)}\begin{pmatrix} \bar{\psi}(r)  \\ \bar{\chi}(r)  \end{pmatrix} \,,
\end{equation}
\noindent the Dirac equation (\ref{2}) (with $k=0$) becomes
\begin{eqnarray}
\hat{O}\bar{\chi}(r)&=&\left[(E-V(r))-\frac{\Omega}{2}\sigma^3-(M+S(r))\right]\bar{\psi}(r),\label{16} \\
\hat{O}\bar{\psi}(r)&=&\left[(E-V(r))-\frac{\Omega}{2}\sigma^3+(M+S(r))\right]\bar{\chi}(r).\label{17}
\end{eqnarray}
\noindent where $\hat{O}=\Omega r(E-V(r))\sigma^2-i\sigma^1 \partial_r + \frac{m_{\alpha}}{r}\sigma^2-\frac{1}{2r}\sigma^{2}\sigma^{3}$ with $m_{\alpha}=m/\alpha$. Using the expression for $\bar{\chi}(r)$ obtained from (\ref{17}) with $V(r)=S(r)$, redefining the spinor as $\bar{\psi}(r)=\frac{\psi(r)}{\sqrt{r}}$ and inserting it in (\ref{16}) we obtain 
\begin{equation}\label{eq2orden2}
\begin{split}
   & \psi^{\prime\prime}(r)+\Bigg[(E^2-M^2)-2V(r)(E+M) \\
   & -2\Omega(E-V(r))\sigma^3+\frac{\Omega^2}{4}-\Omega^2r^2(E-V(r))^2\\
  &-2m_{\alpha}\Omega(E-V(r))+\Omega r \left(\frac{dV(r)}{dr}\right)\sigma^3 \\
  &-\frac{\left(m_{\alpha}-\frac{\sigma^3}{2}\right)^2-\frac{1}{4}}{r^2}\Bigg]\psi(r)=0\,.
\end{split}
\end{equation}
\noindent The Eq. (\ref{eq2orden2}) is effectively a Schr\"{o}dinger-type equation. The second order differential equation obtained in \cite{EPJC79:541:2019} [Eq. (2.18)] is not similar to our result (\ref{eq2orden2}) probably due to erroneous calculations in the manipulation of the two coupled first order differential equations (\ref{16}) and (\ref{17}). 

As in Ref. \cite{EPJC79:541:2019}, firstly we concentrate our efforts for $V(r)=0$. Using $\psi(r)=\begin{pmatrix} \psi_{+}  \\ \psi_{-}  \end{pmatrix}$, $\sigma^{3}\psi_{s}(r)=s\psi_{s}(r)$ with $s=\pm 1$ and $V(r)=0$, (\ref{eq2orden2}) reduce to
\begin{equation}\label{eq2o_0}
\psi_{s}^{\prime\prime}(r)+\left(\lambda_{3}-\lambda_{1}r^{2}-\frac{\lambda_{2}}{r^{2}} \right)\psi_{s}(r)=0\,,
\end{equation}
\noindent where
\begin{eqnarray}
\lambda_{1}&=&E^{2}\Omega^{2}\,,\label{l1}\\
\lambda_{2}&=&\left(m_{\alpha}-\frac{s}{2}\right)^{2}-\frac{1}{4}\,,\label{l2}\\
\lambda_{3}&=&E^{2}-M^{2}-2E\Omega s+\frac{\Omega^{2}}{4}-2m_{\alpha}\Omega E\,.\label{l3}
\end{eqnarray}
\noindent The equation of motion (\ref{eq2o_0}) describes the quantum dynamics of a Dirac particle in a Som-Raychaudhuri space-time. The expression for $\lambda_{2}$ obtained in Ref. \cite{EPJC79:541:2019} [Eq. (2.21)] is wrong. The solution for (\ref{eq2o_0}) with $\lambda_{1}$ and $\lambda_{3}$ real is precisely the well-known solution of the Schr\"{o}dinger equation for the harmonic oscillator. The solution for all $r$ can be expressed as
\begin{equation}\label{solg}
\psi_{s}(r)=N_{n}r^{\left|m_{\alpha}-\frac{s}{2}\right|+\frac{1}{2}}\mathrm{e}^{-\sqrt{\lambda_{1}}r^{2}/2}L^{\left|m_{\alpha}-\frac{s}{2}\right|}_{n}(\sqrt{\lambda_{1}}r^{2})\,,
\end{equation}
\noindent where $N_{n}$ is a normalization constant. Moreover, the spectrum is expressed as (for $E\Omega>0$)
\begin{equation}\label{ener1}
E=\nu_{n,m}+\sqrt{\nu_{n,m}^{2}+M^{2}-\frac{\Omega^{2}}{4}}\,.
\end{equation}
\noindent with $\nu_{n,m}=\left(2n+1+\left|m_{\alpha}-\frac{s}{2}\right|+m_{\alpha}+s\right)\Omega$. The eigenvalue of energy (2.25) obtained in Ref. \cite{EPJC79:541:2019} is not similar to our result (\ref{ener1}) due to it was obtained from a wrong differential equation.

As a second example, let us consider an attractive Coulomb potential $V(r)=-\frac{a}{r}$. By introducing the Coulomb potential into equation (\ref{eq2orden2}), $\psi(r)=\begin{pmatrix} \psi_{+}  \\ \psi_{-}  \end{pmatrix}$ and using , $\sigma^{3}\psi_{s}(r)=s\psi_{s}(r)$ with $s=\pm 1$, we get
\begin{equation}\label{effeq}
\dfrac{d^{2}\psi_{s}}{dr^{2}}+\left[\mathcal{E}^{2}+\dfrac{A}{r}-Br-Cr^{2}-\dfrac{\left(m_{\alpha}-\dfrac{s}{2}\right)^{2}-\dfrac{1}{4}}{r^{2}}\right]\psi_{s}=0\,,
\end{equation}
\noindent where
\begin{eqnarray}
\mathcal{E}^{2}&=&E^{2}-M^{2}-2E\Omega s+\frac{\Omega^{2}}{4}-2m_{\alpha}E\Omega-\Omega^{2}a^{2}\,,\label{eeff}\\
A&=&2\left( E+M \right)a-\Omega as-2m_{\alpha}\Omega a\,,\label{a}\\
B&=&2E\Omega^{2}a\,,\label{b}\\
C&=&E^{2}\Omega^{2}\,.\label{c}
\end{eqnarray}
\noindent The solution for (\ref{effeq}), with $C$ necessarily real and positive, is the solution of the Schr\"{o}dinger equation for the three-dimensional harmonic oscillator plus a Cornell potential \cite{JPA19:3527:1986,RONVEAUX1995,PRC86:052201:2012,EPJC78:494:2018}. By setting
\begin{equation}\label{ansatz}
\psi_{s}=r^{\frac{1}{2}+\left|m_{\alpha}-\frac{s}{2}\right|}\exp \left( -\frac{\sqrt{C}}{2}\,r^{2}-\frac{B}{2\sqrt{C}}%
\,r\right) \phi_{s}(r)
\end{equation}
\noindent and by introducing the new variable and parameters%
\begin{eqnarray}
x &=& \sqrt[4]{C}\,r\,,\label{newv}\\
\omega &=& 2\left| m_{\alpha}-\frac{s}{2} \right|\,,\label{omega}\\
\rho &=& \frac{B}{\sqrt[4]{C^{3}}}\,,\label{rho}\\
\tau &=& \frac{B^{2}+4\,C\,\varepsilon ^{2}}{4\sqrt{C^{3}}}\,,\label{tau}
\end{eqnarray}
one finds that the solution for all $r$ can be expressed as a solution of
the biconfluent Heun differential equation \cite{PRC86:052201:2012,EPJC72:2051:2012,PLA376:2838:2012,AP341:86:2014,AP355:48:2015,EPJC78:44:2018,EPJC78:494:2018}.
\begin{equation}\label{heuneq}
x\,\frac{d^{2}\phi_{s}}{dx^{2}}+(\omega +1-\rho x-2x^{2})\,\frac{d\phi_{s}}{dx}+\left[
\left( \tau -\omega -2\right) x-\Theta \right] \phi_{s}=0\,,
\end{equation}%
\noindent with $\Theta =\frac{1}{2}\left[ \delta +\rho \left( \omega +1\right) \right]\,,$ and
\begin{equation}\label{delta}
\delta=-\frac{2A}{\sqrt[4]{C}}\,.
\end{equation}
\noindent It is known that biconfluent Heun equation has a regular singularity at $x=0$ and an irregular singularity at $x=\infty$ \cite{RONVEAUX1995}. The regular solution at the origin is
\begin{equation}
H_{b}\left(\omega,\rho,\tau,\delta;x\right)=\sum_{j=0}^{\infty}\frac{1}{\Gamma(1+j)}
\frac{A_{j}}{j!}x^{j}\,,
\end{equation}
\noindent where $\Gamma(z)$ is the gamma function, $A_{0}=1$, $A_{1}=\Theta$ and the remaining coefficients for $\rho\neq 0$ satisfy the recurrence relation,
\begin{equation}\label{recu}
\begin{split}
A_{j+2}= & \left[ (j+1)\rho+\Theta \right]A_{j+1}\\
&-(j+1)(j+\omega+1)(\Delta-2j)A_{j}\,,
\end{split}
\end{equation}
\noindent where $\Delta=\tau-\omega-2$. The series is convergent and tends to $\exp\left( x^2+\rho x \right)=\exp\left(  \sqrt{C}r^{2}+\frac{B}{\sqrt{C}}r\right)$ as $x\rightarrow\infty$. This asymptotic behavior perverts the normalizability of the solution (\ref{ansatz}), because $\psi(r)\propto \exp\left(  \frac{\sqrt{C}}{2}r^{2}+\frac{B}{2\sqrt{C}}r\right)$ as $r\rightarrow\infty$. This impasse can be surpassed by considering a polynomial solution for $H_{b}$. From the recurrence (\ref{recu}), $H_{b}$ becomes a polynomial of degree $n$ if only if two conditions are satisfied:
\begin{equation}\label{c1}
\Delta=2n, \qquad (n=0,1,2,\ldots)
\end{equation}
\noindent and
\begin{equation}\label{c2}
A_{n+1}=0\,.
\end{equation}
\noindent The condition (\ref{c2}) furnishes a polynomial of degree $n+1$ in $\delta$, there are at most $n+1$ suitable values of $\delta$. At this stage, it is worth to mention that the energy of the system is obtained using both conditions (\ref{c1}) and (\ref{c2}).

From the condition (\ref{c1}), we obtain (for $E\Omega>0$)
\begin{equation}\label{enm}
E_{n,m}^{2}-M^{2}+\frac{\Omega_{n,m}^{2}}{4}-2\Omega_{n,m}\xi_{n,m}E=0\,,
\end{equation}
\noindent where
\begin{equation}\label{xi}
\xi_{n,m}=s+m_{\alpha}+\left|m_{\alpha}-\frac{s}{2}\right|+n+1\,.
\end{equation}
\noindent The problem does not end here, it is necessary to analyze the condition (\ref{c2}). For $n=0$, the condition (\ref{c2}) becomes $A_{1}=\Theta=0$ and results in an algebraic equation of degree one in $\delta$
\begin{equation}\label{c2n0}
\delta+\rho(\omega+1)=0\,.
\end{equation}
\noindent Substituting (\ref{omega}), (\ref{rho}) and (\ref{delta}) into (\ref{c2n0}), we obtain
\begin{equation}\label{omegan0m}
\Omega_{0,m}=\frac{2\left(E+M\right)}{\epsilon_{m}}\,,
\end{equation}
\noindent where $\epsilon_{m}=2\left(m_{\alpha}+\frac{s}{2}\right)+2\left|m_{\alpha}-\frac{s}{2}\right|+1$. Substituting (\ref{omegan0m}) into (\ref{enm}) for $n=0$, we have
\begin{equation}\label{en0m}
E_{0,m}=M\frac{\epsilon_{m}^{2}-1}{1-\epsilon_{m}^{2}-2(s+1)\epsilon_{m}}\,.
\end{equation}
\noindent The expression (\ref{en0m}) represents the energy eigenvalue for $n=0$. For $n=1$, the condition (\ref{c2}) becomes $A_{2}=0$ and results in an algebraic equation of degree two in $\delta$. For $n\geq 2$ the algebraic equations are cumbersome. In this comment, we will only consider the solution for $n=0$ for simplicity.
 
In summary, we studied the Dirac equation in presence of scalar and vector potentials in a class of flat G\"{o}del-type space-time called Som-Raychaudhuri space-time. We calculated the correct second order differential equation for this system. As in Ref. \cite{EPJC79:541:2019}, we have considered two case: ($1$) $V(r)=0$ and ($2$) Coulomb potential. For the first case $V(r)=0$, the problem was mapped into a Schr\"{o}dinger-like equation with the harmonic oscillator potential. The correct energy spectrum for this case was obtained. For the second case, we considered an attractive Coulomb potential. In this case, the problem was mapped into biconfluent Heun differential equation and using appropriately the quantization conditions (\ref{c1}) and (\ref{c2}), we found the correct energy spectrum for $n=0$. Finally, we showed that the results obtained in Ref. \cite{EPJC79:541:2019} are incorrect, due to they were obtained from a wrong differential equation. 

\begin{acknowledgements}
This work was supported in part by means of funds provided by CNPq, Brazil, Grant No. 307932/2017-6 (PQ) and No. 422755/2018-4 (UNIVERSAL), S\~{a}o Paulo Research Foundation (FAPESP), Grant No. 2018/20577-4, FAPEMA, Brazil, Grant No. UNIVERSAL-01220/18 and CAPES, Brazil.
\end{acknowledgements}


\end{document}